Phosphorene as an Anode Material for High Performance Lithium-Ion Battery: First Principle Study and Experimental Measurement

Congyan Zhang, George Anderson, Ruchira Ravinath Dharmasena, Gamini Sumanasekera, and Ming Yu*

Department of Physics and Astronomy, University of Louisville, Louisville, KY 40292

**Abstract**: The prospects of phosphorene as an anode material for high performance Li-ion battery was systematically investigated from the first principle calculations and experimental measurements. The diffusion energy barriers of a Li atom moving along various orientations on phosphorene layer were calculated from the Li adsorption energy landscape. It was found that the diffusion mobility of a Li atom along the zigzag direction in the valley of phosphorene could be about 7 to 11 orders of magnitude faster than that along the other directions, indicating its ultrafast and anisotropic diffusivity. The lithium insertion in phosphorene was studied considering various $Li_nP_{16}$ configurations ($n$=1~16). It was found that phosphorene could accommodate up to one Li per P atom (*i.e.*, $Li_{16}P_{16}$), and the predicted theoretical value of the Li capacity for a single layered phosphorene can reach about 865 *mAh/g*. Our experimental measurement on the Li capacity for a network of a few layered phosphorene can reach a reversible stable value of ~ 453 *mAh/g* even after 50 cycles. In particular, it was found that, even at the high Li concentration (*e.g.*, $x = 1$ in $Li_xP$), there was no Li clustering and the structure of phosphorene is reversible during the lithium intercalation. Our results clearly show that phosphorene has promise as a novel anode material for high performance Li-ion batteries.

*Corresponding author's email address: m0yu0001@louisville.edu

1. Introduction

One of the major challenges in developing anode materials for high performance Lithium-ion battery (LIB) is to find promising anode materials with high capacity, high rate charging/discharging, large open circuit voltage, and good reversibility in cycling performance.[1,2] Graphite has conventionally been chosen as the anode in LIB for commercial use because of its high energy stability, low cost, and good cycling performance.[3-7] But its low specific capacity of 372 *mAh/g* (*i.e.*, its maximum limit to intercalate Li is one Li atom per 6 carbon atoms under ambient condition[8,9]) limited its application as a promising anode for high performance LIB, and demands a search for better performing anode martials beyond graphite. Silicon was found to be a very attractive anode with extremely high capacity of 4200 *mAh/g*.[10-15] However, the large volume expansion, due to the formation of the $Li_{22}Si_5$ alloy during lithium intercalation, leads to an irreversibility of Li insertion/extraction[16], and therefore its stability in cycling performance is problematic. Intensive studies on nanostructured Si anode materials to enhance the cycling life have been performed.[17-19] Recently, black phosphorus has aroused much interest as promising anode material in LIB because of its weak van der Waals interlayer space and relatively high capacity of 2596 *mAh/g*.[20-25] Unfortunately, the cycling efficiency is limited due to the large volume expansion (~ 300 % expansion[21]) when Li atoms insert into the black phosphorus causing the combined system to form $Li_3P$ alloy.

Quite recently, two-dimensional (2D) materials have been considered as potential anode materials for high performance LIB because of their high charge carrier mobility[26], large surface-volume ratio[27], and broad electrochemical window[28], as compared to their bulk counterparts, which enable fast ion diffusion and offer more ion insertion channels with the entire surface exposed. Two typical promising 2D anode materials in LIB are graphene[29-47] and transition metal dichalcogenides (TMDCs), such as Molybdenum disulfide ($MoS_2$).[48-56] The theoretical capacity of graphene was predicted to be 744 *mAh/g* with the hypothesis that Li atoms could be adsorbed on each side of the graphene with the $LiC_3$ configuration, and the separation of Li atoms could still keep about 4.35 Å to avoid Li clustering, as found in the graphite.[42-44] However, the experimental reports showed that at ambient temperature and pressure, Li metal invariably co-exist with the parallel layers of graphene[42], and only the crumpled or curved singe layered graphene nanosheet[44,45] or carbonaceous such as a group of single layered graphene arranged in 'house card'[43] could accommodate up to two Li atoms per six carbon atoms on both sides of the graphene as well as on the edges. The reversible capacity of such curved graphene was found depending on the nanoporosity of the graphene.[42-45] Similar results were also found in graphene-like $MoS_2$.[57] Furthermore, it was pointed out that the weak interactions between Li and the graphene (*e.g.*, the binding energy is about 1.04 eV[46]) might result in the low open circuit voltage and weaken the electrochemical performance[7, 28-31] in LIB. Searching for anode materials with strong interaction with Li atoms and large open circuit voltage is motivated.

Analogous to graphene and MoS$_2$, phosphorene can be synthetized by mechanical exfoliation of the black phosphorus.[58,59] This new discovered 2D material has quite unique properties. It exhibits a puckered honeycomb structure, and possesses semiconductor behavior with direct band gap and the anisotropic electronic properties. In addition to those properties, the anisotropic mechanical, thermal and optical properties and the related potential applications have been reported.[60-67] Based on the unique properties, numbers of theoretical calculations about the diffusion and the capacity of phosphorene in Li-ion and Na-ion batteries have been carried out.[68-76] Two theoretical groups found that the diffusion energy barrier of Li in phosphorene is anisotropic[68,69], and the diffusion mobility of Li atom along the zigzag direction on phosphorene is about $10^{10}$ times faster than that along the armchair direction. The similar ultrafast diffusion mobility was also reported in Na-ion battery.[74,75] Noted that these calculations on the diffusion were carried out using the climbing image nudged elastic band method[77,78] along some specific orientations, and a comprehensive study on the diffusivity along various possible diffusion paths is desired so as to have an entire picture about the diffusivity on phosphorene surface. The theoretical capacity of Li on the phosphorene monolayer was predicted to be about 433 *mAh/g* by Zhao *et al*. using DFT calculations.[73] While Kulish *et al*.[75] predicted that the capacity for Na can be as high as 865 *mAh/g* using DFT calculations including the van der Waals interaction.[79] Therefore, a complete understanding on the capacity of Li in phosphorene is indeed necessary, and experimental investigation along this orientation is indispensable.

In this work, we performed a systematically study on the Li adsorption energy landscape, the diffusion process, the ability of phosphorene to accommodate Li atoms, and the capacity of Li in phosphorene using the first principle calculations. The interaction between Li and phosphorene was found much stronger than that in the graphene implying the possible high open circuit voltage and potential application to enhance the electrochemical performance. In particular, we found that the Li diffusion on the phosphorene is about $10^4$ and $10^2$ times faster than on the graphene and $MoS_2$, which implies that phosphorene may exhibit outstanding rate charging/discharging. Estimated theoretical Li capacity can reach as high as 865.5 *mAh/g*, similar as the value that predicted by Kulish *et al*. for Na in phosphorene.[75] Our experimental measurements on the Li capacity for a few layered phosphorene networks show a reversible and stable capacity of ~ 453 *mAh/g* after 30 cycles. Such value is remained for over 50 cycles. Interestingly, no Li clustering was found, and even if some of the phosphorous bonds broke at a high Li concentration, they could automatically reform after removing Li atoms, indicating the good reversibility of phosphorene during the insertion/extraction process. Our current results clearly reveal the outstanding feature of phosphorene as promising host materials for high performance LIB.

2. **Computational methods**

Our first principles calculations were performed using the density functional theory (DFT)[80,81] framework, as implemented in the Vienna Ab-initio Simulation Package (VASP).[82] The electron-ion interactions were described by Projector Augmented Wave (PAW)[83], while electron exchange-correlation interactions were described by the generalized gradient approximation (GGA)[84] in the scheme of Perdew Burke Ernzerhof (PBE). The van der Waals interaction was not included in our calculation because it was pointed out that, instead of Na, the van der Waals interaction plays a minor role for a Li adsorption on the phosphorene.[73] An energy cutoff was set to be 500 eV for the plane wave basis in all calculations, and the criteria for the convergence of energy and force in relaxation processes were set to be $10^{-4}$ eV and $10^{-4}$ eV/Å, respectively.

The optimized lattice constants for phosphorene are $a$ = 3.306 Å, and $b$ = 4.619 Å, which are consistent with the previous DFT results.[58,69,73-74] The band structure for the optimized phosphorene shows the directed bandgap behaver with the band gap of 0.82 eV, in consistent with other DFT results[58,73,75], but underestimated compared with the experimental results (*e.g.*, 1.45 eV reported in Ref. 58, or 2.05 eV reported in Ref. 85, or 2.09 eV reported in our present work). A 2×2 supercell was chosen to study the adsorption, the diffusion, and the capacity of Li on phosphorene monolayer. A vacuum space of 15 Å was set between adjacent layers to avoid any mirror interactions of Li atoms. The Brillouin zone is set using a Monkhorst-Pack grid of 16×12×1 in all calculations.

The adsorption energy per Li atom on monolayer phosphorene ($E_a$) is defines as

$$E_a = (E_{Li_nP_{16}} - E_{P_{16}} - nE_{Li})/n, \tag{1}$$

where $E_{Li_nP_{16}}$, $E_{P_{16}}$, and $E_{Li}$ are the total energy of the $Li_nP_{16}$ system, the 2×2 pristine phosphorene (16 atoms), and the isolated Li atom, respectively, and $n$ is the number of adsorbed Li atoms in the 2×2 supercell. From the energy adsorption defined in Eq. (1), we can calculate the adsorption energy landscape for a single Li atom loaded on phosphorene surface and find the preferential positions for Li adsorbed on the phosphorene. The open circuit voltage (OCV) used in the battery fields can be also interpreted in terms of the formation energy involved in the adsorption process since the entropy and the pressure contributions to the Gibbs free energy are negligible.[86] If we take lithium metal as the cathode, the OCV is then related to the adsorption energy (defined in Eq. (1)) by $\bar{V} \propto \frac{E_a}{zF}$, where $F$ is the Faraday constant (96,485 *C/mol*), and $z$ is the charge (in electron) transferred by Li in the electrolyte and can be approximately taken as one for Li in the case of the atomic ratio of Li to P to be 1:1.

Based on the adsorption energy landscape, the energy barrier along different diffusion paths was determined. Following the Arrhenius equation[87] for the temperature dependence of reaction rates, and currently can be used to model the temperature variation of diffusion coefficients, the diffusion constant ($D$) of Li can be estimated as

$$D \sim \exp\left(\frac{E_b}{kT}\right), \qquad (2)$$

where $E_b$ is the activation energy (or diffusion barrier), $k$ is the Boltzmann constant, and $T$ is the environmental temperature (300 K was chosen in the calculation). Following Eq. (2), the diffusion constant of Li on phosphorene can be qualitatively evaluated, and the most preferential diffusion path for single Li atoms on phosphorene can be determined.

The theoretical capacity of LIB was estimated by Faraday's law through studying the adsorption energy for Li atoms adsorbed on phosphorene. It is defined as

$$C = \frac{\lambda F}{M_P}, \qquad (3)$$

where $\lambda$ is the atomic ratio of Li to P atoms, $F$ is the Faraday constant (26.8 *Ah/mol*), and $M_P$ is the atomic mass of P atom (31 *g/mol*), respectively.

3. **Experimental methods**

3.1 **Synthesis**

Bulk black phosphorus was first synthesized from red phosphorous by using a short transport growth method.[88] Phosphorene was then synthesized from the bulk black phosphorus by using liquid mechanical exfoliation.[89] Bulk black phosphorus (50mg) was suspended in Dimethyl Formaldehyde (DMF) (which was chosen to prevent the interaction of phosphorene with water). The mixture was sonicated up to 1 hour and 30 minutes. Samples were then centrifuged at 25 $^{o}$C for 30 minutes at 10,000 rpm. Centrifugation separated into two distinct regions that contained either 1-5 layers (referred as the top centrifuge in Fig. 1) or 5-13 layers phosphorene (referred as the bottom centrifuge in Fig. 1). Then the solution containing 1-5 layers was evaporated slowly inside an Argon glove box and the resulting powder was scraped off the container. Sample exposure to air was kept to a minimum during characterization to prevent undesired reactions. Raman Spectroscopy and photoluminescence were done using Renishaw inVia Raman Microscope. For Raman Spectroscopy, the excitation laser wavelength was 632 nm at 10% laser power. Photoluminescence was done using 442 nm wavelength excitation laser at 10%. Transmission Electron Microscopy/Energy Dispersive X-Ray Spectroscopy/Electron Diffraction was done on films supported by lacey carbon on copper grids. Atomic Force Microscopy was done using Asylum MF3D AFM. Scanning Electron Microscopy was done using Nova FEG-SEM.

### 3.2 Characterization

In Fig. 1 (a), the Raman spectrum of the samples is shown. The characteristic peaks of phosphorene are $A_{1g}$, $B_{2g}$, $A_{2g}$ and have been calculated to be at 368 cm$^{-1}$, 433 cm$^{-1}$, and 456 cm$^{-1}$ respectively.[90] The current work found 362.34 cm$^{-1}$, 439 cm$^{-1}$, 467.56 cm$^{-1}$ for samples taken from 1-5 layer phosphorene (*i.e.*, the top centrifuge). It has been found that the shifts in peak position for phosphorene are too small to be used to determine layering number. However, previous studies have found that the intensity of the $A_{1g}$ peak in relation to the Si peak can be used to determine the number of layers of phosphorene.[91] Based on the above intensity ratio calibration, the current Raman spectrums show the number of layers to be between 3 and 9 layers. Raman spectrum from the precursor material (red phosphorus) was included to show that no red phosphorus peaks are seen in the phosphorene samples. Photoluminescence (Fig.1 (b)) shows that for bulk black phosphorus no peaks are present in the region of interest due to black phosphorus having a band gap of 0.3 eV.[92] The samples that had the smallest layering number (taken from top of centrifuge vial, and represented by red curve in Fig. 1 (b)) have a strong peak at 593.28 nm (2.09 eV) and weaker peaks at 696.00 nm (1.78 eV) and 769.00 nm (1.61 eV). The weaker peaks are believed to be due to none uniform distribution of layering number across the substrate and has been found to be a common issue with liquid mechanical exfoliated samples.[89] The strongest peak at 2.09 eV, is suspected to be due to monolayer phosphorene. Photoluminescence can be approximated as the lower bound of the energy band gap, due to the emission being caused by exciton recombination. Energy band gaps of 1.6 eV[91] and 1.29 eV[93] have been found for two layers phosphorene from photoluminescence estimates, which is smaller than the value (2.09 eV) found in the current work for single layer phosphorene.

Atomic Force Microscopy (AFM) is an excellent method for determining the number of atomic layers in 2D materials. It has been found that 0.9 nm is one layer of phosphorene.[58] Fig. 1 (c) shows that the current work has a step height of 0.99 nm for a phosphorene flake and is consistent with one monolayer. Scanning Electron Microscopy (SEM) in Fig. 1 (d) confirms the non-uniform distribution of layering that the photoluminescence data showed. SEM shows overlapping flakes, which would give multiple photoluminescence signals as previously found. The uniformity of the layering number may be improved by using spin coating.[89]

The electrochemical characterizations were carried out using a CR2032 coin cell-type cell using an Arbin battery tester. The anode electrode was formed by casting a mixture of phosphorene and poly tetrafluroethylene (PTFE) coated teflonized acetylene black (TAB-2) and then pressing the mixture onto a stainless steel mesh. A lithium foil with area of 1.2 cm$^2$ was used as a cathode and was separated by porous glass fiber. The electrolyte was LiPF6-ethylene carbonate (EC)-diethyl carbonate (DEC) [1 M LiPF6-EC:DEC (1:1) by volume]. The charge/discharge performance was carried out in the voltage range of 0.1-2.8 V at a C/10 rate.

## 4. Results and discussion

### 4.1 Adsorption energy landscape

The adsorption energy landscape of a Li atom adsorbed on phosphorene was obtained by performing the vertical relaxation of the Li-P system with a single Li atom loading at different sites above phosphorene. 40 possible Li adsorption positions were considered in the 1x1 unit cell of the phosphorene as shown in Fig. 2 (a), in which the Li atom was placed either on the top of P atoms (referred as TA sites, and are represented by red dots), or on the top of the middle of P-P bonds (referred as TB and VB sites, and are represented by blue dots), or at the center of the triangular region formed by P-P-P atoms (referred as TH and VH sites, and are represented by green dots), or at other sites spread among these sites (represented by black dots). The characters 'V' and 'T' in the notations represent the positions of Li atom in the valley and on the top of ridge, respectively. The vertical relaxation was carried out by allowing the motion of Li and P atoms along the direction perpendicular to the phosphorene surface. The vertical distance ($d$) which is defined as the distance between the Li atom and the middle of the thickness of puckered phosphorene (as shown in Fig. 2 (b)) was optimized at different adsorption sites. Fig. 2 (c) shows one example of the total energy as a function of the vertical distance $d$ for a Li atom loaded at the VH site. The optimized vertical distance at VH site is 2.53 Å.

Based on the calculated adsorption energy on each site, the adsorption energy landscape was obtained (see Fig. 3). The colors in the right column in Fig. 3 indicate the adsorption energies ($E_a$) at different sites relative to that at site VH. The darker the color is, the lower the relative adsorption energy. Fig. 3 clearly shows that the adsorption energy landscape possesses an anisotropic behavior. Positions of the Li atom with relatively low adsorption energy are located in the valley along the zigzag direction with minimum energy at VH sites, while positions of the Li atom with relatively high energy are located on the top of the ridge along the zigzag direction with the maximum relative energy at TA sites. The adsorption energy difference between these two sites is about 0.78 eV/Li. The saddle points were found at the VB site and the TB site, as indicated by the white and blue boxes in Fig. 3 (a). Their corresponding adsorption energies are lower than those of neighbors along the bond direction, but higher than those of the neighbors along direction perpendicular to the bond, as shown in details in Fig. 3 (b) and (c). The adsorption energy landscape reveals that Li prefers to stay at the most stable VH site, also possibly stay at the metastable VB and TB sites when the most stable VH sites are occupied. Furthermore, the energy difference along the zigzag direction in the valley is much smaller than that along the armchair direction, which indicates that the preferential diffusion path for single Li atom will be along the zigzag direction in the valley.

### 4.2 Diffusion

Rate charging and discharging in the LIB relates to the mobility of Li ions in the anode/cathode. The faster the Li atom moves, the higher the charging/discharging rate, and therefore the rate capacity in the LIB. To study how fast the Li atom moves/diffuses on the phosphorene, we calculated the diffusion energy barrier $E_b$ (defined by the relative adsorption energy at site along the diffusion pathway with respect to that at the corresponding initial site) along various diffusion paths between the preferential adsorption sites VH, VB and TB. Fig. 4 presented the most four possible diffusion paths: in path 1, the Li atom migrates along the zigzag direction, starting at a VH site, going through VB sites, and then ending to another VH site (referred as VH→VB→VH, see the black dots in Fig. 4 (a)); in path 2, Li atom migrates along the direction perpendicular to the zigzag direction, starting at a VB site, crossing over the ridge through TB sites, and then ending to another VB site (referred as VB→TB→VB, see the green dots in Fig. 4 (a)); in path 3, Li atom migrates along the direction that starts at a VH site, crosses over the ridge through TB sites, and then ends to another VH site (referred as VH→TB→VH, see the blue dots in Fig. 4 (a)); and in path 4, Li atom migrates along the armchair direction, starting at a VH site, crossing over the ridge through the highest adsorption energy sites TA, and then ending to another VH site (referred as VH→TA→VH see the red dots in Fig. 4 (a)). The two different side views of these four paths were also shown in Fig. 4 (b) and (c). Cleary, Li atom migrates in the same valley along path 1, but it migrates from one valley over the ridge and to another valley along paths 2-4.

The corresponding energy barriers ($E_b$) with respect to the initial position energy for each path as function of the relative distance are summarized in Fig. 5 (a). The relative distance of Li atom is defined by the ratio of horizontal distance of Li with respect to the starting point to the horizontal distance of the ending point with respect to the starting point. It was found that the adsorption energy at the VB site is relatively higher than that at other sites along the path 1, and the calculated corresponding diffusion energy barrier to the VH site is 0.09 eV. In path 2, the corresponding diffusion energy barrier is about 0.58 eV, which is about 0.49 eV higher than that in the path 1. It is found that there is a very shallow valley in the diffusion energy barrier around the TB site in path 2, which is attributed to the local minimum along the direction of the P-P bond at the TB site (see Fig. 4 (c)). Similarly, Li moving along the path 3 also crosses the TB site with the initial point at the VH site, but since the path 3 is along the direction perpendicular to the P-P band at the site TB, there is no local minimum around the TB site. The calculated corresponding diffusion energy barrier in path 3 is about 0.69 eV, which is about 0.11 eV higher than that in the path 2 and about 0.60 eV higher than that in the path 1. In path 4, Li atom migrates from the most preferential adsorption site VH along the armchair direction, and crosses over the most unstable site TA; the calculated diffusion energy barrier is 0.78 eV which is the highest one among these diffusion paths. The Li diffusion constant can be quantitatively evaluated from Eq. (2). The evaluated value along the path 1 is about $3.58 \times 10^7$ times faster than that along the path 2, about $7.36 \times 10^9$ times faster than that along the path 3, and about $1.89 \times 10^{11}$ times faster than that along the path 4, respectively, indicating that the Li diffusion on phosphorene is ultrafast and anisotropic. In addition, the comparison of diffusion constants among these paths reveals that even though paths 2-4 are all cross over the ridge, the mobility of Li is quite different. It is also noted that the diffusion energy barrier of Li on phosphorene (0.09 eV) is much lower than that on graphene (0.327 eV)[47] and $MoS_2$ (0.25 eV).[57] The evaluated corresponding mobility of Li on phosphorene is about 4 and 2 order of magnitude faster than that on graphene and $MoS_2$, respectively, indicating that the rate charging/discharging is much higher in phosphorene than in graphene and $MoS_2$. Such high Li diffusivity is important to satisfy the current-density requirements and is essential for the performance of the anode materials in LIB.

The ultrafast and anisotropic diffusion nature for a Li migrates on the phosphorene comes from the unique puckered structure of phosphorene, which is crucial important for the Li diffusion. Fig. 5 (b) shows the optimized vertical distances *d* as function of the relative distance along each path. It was found that the maximum vertical difference along the path 1 is 0.09 Å, and such small height difference allows Li to migrate though the channel very easily. On the other hand, in paths 2-4, a Li atom has to climb from the valley up to the ridge by about 0.58-0.79 Å, hence much more energy is needed for Li to conquer the ridge. A comparison between the diffusion energy barrier (Fig. 5 (a)) and the vertical distance (Fig. 5 (b)) reveals that the lower the vertical distance *d*, the lower the diffusion energy barrier and the easier for Li to diffuse.

### 4.3 Lithium insertion in $Li_nP_{16}$ system

To study how much phosphorene can accommodate Li atoms and, therefore, to evaluate the capacity of Li on phosphorene monolayer, several $Li_nP_{16}$ systems including $LiP_{16}$, $Li_2P_{16}$, $Li_4P_{16}$, $Li_8P_{16}$, and $Li_{16}P_{16}$ were studied. For each $Li_nP_{16}$ system, several configurations including Li atoms loaded on single side and double sides of the phosphorene were taken into consideration. In all configurations, Li atoms are initially placed in VH, VB or TB sites and the corresponding systems were fully relaxed to obtain the stable configurations with different Li concentration. The results are discussed as follows.

### 4.3.1 LiP$_{16}$ system

The top and side views of the stable LiP$_{16}$ structures with a single Li atom adsorbed at VH, VB, and TB sites are shown in Fig. 6. The corresponding adsorption energy $E_a$ and the geometric properties are summarized in Table 1. It is obviously seen from the 3$^{rd}$ column that the VH site is the most energetically stable adsorption site with the adsorption energy of -2.086 eV/Li. The next stable adsorption site is the VB site (-1.995 eV/Li), following by the TB site (-1.427 eV/Li). From the 4$^{th}$ and 5$^{th}$ columns in Table 1, we found that Li atom located at the VH site with a vertical distance $d$ of 2.53 Å has three nearest P neighbors; the corresponding Li-P distances are 2.45, 2.54 and 2.54 Å respectively. Li atom at the VB site has two nearest P neighbors with Li-P distance of 2.41 Å, and the vertical distance is about 0.07 Å higher than that at the VH site. While at the TB site, the vertical distance is about 0.7 Å higher than that at the VH site, and the two nearest Li-P distances are equal with the value of 2.58 Å. Since the Li-P distances in these configurations are within the Li-P bond length of 2.68 Å, it indicates that Li atom is chemically bonded with P atoms. Such chemical bonding nature and low adsorption energy demonstrate that the interaction between Li and phosphorene is strong, compared to the graphene, and therefore could prevent Li clustering during Li insertion, which is desired in the electrochemical performance.

### 4.3.2 Li$_2$P$_{16}$ system

We considered about 45 initial configurations for 2 Li atoms loaded on the single-side of phosphorene. These single-side configurations can be classified by 6 groups denoted as S-VH-VH, S-VH-VB, S-VH-TB, S-VB-VB, S-TB-TB, and S-VB-TB, respectively, where the first notation 'S' means the single-side adsorption, and the second/third notations indicate the locations of two Li atoms, respectively. We found that only 12 stable structures corresponding to groups of S-VH-VH, S-VH-VB, S-VB-VB, and S-TB-TB were obtained after fully relaxation. In particular, initial configurations belong to groups S-VH-TB and S-VB-TB were not stable and transferred to the configurations belong to the group S-VH-VH after the full relaxation, mostly due to the migration of the Li atom at the VB or TB site to the more stable VH site. Fig. 7 (a) presents the top and side views of the four most stable single-side configurations corresponding to each of the four groups (*i.e.*, S-VH-VH, S-VH-VB, S-VB-VB, and S-TB-TB, respectively). The corresponding adsorption energies and geometric properties for these four configurations are listed in Table 2.

From the adsorption energy $E_a$ (the 3$^{rd}$ column in Table 2), we found that the configuration in the group S-VH-VH (*i.e.*, the first panel in Fig. 7 (a)) is the most stable configuration compared with other single-side configurations (*e.g.*, the 2$^{nd}$~ 4$^{th}$ panels in Fig. 7 (a)). There is no significate change in the vertical distance *d* at sites VH, VB, and TB, comparing with LiP$_{16}$ system when the Li atoms stay on single-side (see the 5$^{th}$ column in Tables 1 and 2). But some of the Li-P distances ($d_{\text{Li-P}}$) decrease in the Li$_2$P$_{16}$ system, for instance, from 2.45 Å to 2.35 Å at VH sites, from 2.41 Å to 2.30 Å at VB site, and from 2.58 Å to 2.45 Å at TB sites, respectively (see the 4$^{th}$ column in Tables 1 and 2), indicating that the more the Li atoms adsorbed, the strong the Li-P attractive interactions, and the shorter the Li-P bonds. In addition, since the Li concentration is still low, the nearest Li-Li distance is large (4.62-4.99 Å, see the 6$^{th}$ column in Table 2) which is longer than the Li-Li bond length (2.68 Å in our DFT calculation), indicating that the interaction between Li atoms is very weak.

By adding one more Li atom on the other side of the three stable configurations obtained in LiP$_{16}$ system, we constructed double-side configurations for Li$_2$P$_{16}$ system. It should be noted that, on the other side of phosphorene, the corresponding stable adsorption sites (*i.e.*, the VH, VB or TB site) are those located below TH, TB, and VB sites, if we see them from the top view of phosphorene (Fig. 2 (a)). 48 initial double-side configurations were constructed following this consideration. They are classified by 6 groups, denoted as D-VH/VH, D-VH/VB, D-VH/TB, D-VB/VB, D-VB/TB, and D-TB/TB, respectively. The notation 'D' indicates the double-side adsorption; the notation before/after the slant indicates the Li atoms at the sites above/below phosphorene monolayer. After fully relax, only 15 stable double-side configurations corresponding to the first five groups were obtained, and the five most stable double-side configurations among them corresponding to each of the five groups are shown in Fig. 7 (b). The corresponding adsorption energies and geometric properties are listed in Table 2.

Comparing the adsorption energies $E_a$ among those stable single/double-side configurations (see the 3$^{rd}$ column in Table 2) we found that the configuration D-VH/VH (-2.113 eV/Li) is the most stable structure in the Li$_2$P$_{16}$ system, followed by configuration D-VH/VB (-2.064 eV/Li), and then the single-side configuration S-VH-VH which is about 0.02 eV higher than the configuration D-VH/VB. It was also found that, by adding one more Li on the other side of the stable LiP$_{16}$ system, the nearest Li-P distance ($d_{Li-P}$) (the 4$^{th}$ column in Table 2) and the vertical distance $d$ (the 5$^{th}$ column in Table 2) at sites VH, VB, and TB are similar to those in the stable LiP$_{16}$ system (the 4$^{th}$ and 5$^{th}$ columns in Table 1), since the separation of Li atoms on both sides are still the same to those in the stable LiP$_{16}$ system.

### 4.3.3 Li$_4$P$_{16}$ system

In searching the stable configurations for Li$_4$P$_{16}$ system with single-side adsorption, we mainly focused on the configurations in which four Li atoms were loaded at each of the four 1x1 unit cells above phosphorene monolayer. The initial single-side configurations were classified by three groups, denoted as S-4VH, S-4VB, and S-4TB, respectively, where the number in the notations indicates the total number of Li atoms on the corresponding sites. These single-side configurations were fully relaxed and only five of them were stabilized to the structures named as S-4VH$_a$, S-4VH$_b$, S-2(VH-TH), S-4VB, and S-4TB, respectively (see Fig. 8 (a)), in which the configurations S-4VH$_a$ and S-4VH$_b$ are distinguished by the distribution of the four Li atoms at the four VH sites on the 2x2 supercell with different symmetry (see the two left panels in Fig. 8 (a)). While, the configuration S-2(VH-TH) (*i.e.*, the second right panel in Fig. 8 (a)) was obtained from the initial configuration in the group S-4VH by distributing the four Li atoms at the VH sites with Li-Li distance shorter than the Li-Li bond length. During the relaxation, these Li atoms push each other away, and two of them move to the TH sites, and the system was finally relaxed to a stable structure with S-2(VH-TH) configuration with Li-Li distances of 3.0 Å.

The corresponding adsorption energies and geometric properties are listed in Table 3. It can be seen from the adsorption energy $E_a$ (the 3$^{rd}$ column) that configurations S-4VH$_a$ and S-4VH$_b$ are two relatively stable structures, as compared to the other three stable single-side configurations. Comparing the values in the 4$^{th}$ and 5$^{th}$ columns in Tables 1 and 3, we found that the nearest Li-P distances keep almost the same values in both $LiP_{16}$ and $Li_4P_{16}$ systems, while the vertical distances $d$ of the Li atom at VH, VB and TB sites in $Li_4P_{16}$ are about 0.1-0.3 Å higher than those in $LiP_{16}$, indicating a vertical increase of phosphorene layer when more Li atoms attracted by P atoms, and therefore, a slightly expansion in volume. This phenomenon is also found when more Li atoms adsorbed to phosphorene (see the results and discussions for $Li_8P_{16}$ and $Li_{16}P_{16}$ systems in the following sections).

Similar to our consideration in searching stable double-side configurations for $Li_2P_{16}$ structures, we added two more Li atoms on the other side of the four stable single-side configurations in $Li_2P_{16}$ system (*i.e.*, the four structures shown in Fig. 7 (a)), and constructed 36 initial double-side configurations. Among them, only four stable structures are obtained after fully relax. They are denoted by D-2(VH/VH)$_{a/b}$, D-2(VB/VB), and D-(VH-VB/VH-VB), respectively, where the subscripts a and b indicate the distribution of the four Li atoms in the same configuration with different symmetry. The top and side views of these stable structures are shown in Fig. 8 (b) and the corresponding energetic and geometric properties are also listed in Table 3. Comparing the adsorption energy $E_a$ (the 3$^{rd}$ column in Table 3) for all the obtained stable configurations in $Li_4P_{16}$ system, we found that all the double-side configurations have lower in energy than those of single-side configurations, and the relatively more stable configurations are the double-side configurations D-2(VH/VH)$_a$ and D-2(VB/VB), followed by the double-side configuration D-(VH-VB/VH-VB). Furthermore, as shown in the last column in Table 3, the Li-Li distance in these relative stable structures are 4.62 Å in configurations D-2(VH/VH)$_a$ and D-2(VB/VB) and 4.40 Å in the configuration D-(VH-VB/VH-VB), therefore, the repulsive interactions between them are still weak

### 4.3.4 $Li_8P_{16}$ system

When two Li atoms were inserted to each 1x1 unit cell at VH, VB and TB sites forming VH-VH, VH-VB, VH-TB, VB-VB, VB-TB, and TB-TB pairs, the corresponding initial single-side configurations were constructed. They are denoted by S-4(VH-VH), S-4(VH-VB)$_{a/b}$, S-4(VH-TB)$_{a/b}$, S-4(VB-VB), S-4(VB-TB)$_{a/b}$, and S-4(TB-TB), respectively, where the subscripts a and b indicate the distribution of the eight Li atoms in the same configuration with different symmetry. These structures were fully relaxed and only five of them were found stable. Fig. 9 (a) shows the top and side views of the five stable $Li_8P_{16}$ structures with Li atoms located at the single-side adsorption. The corresponding adsorption energies and geometric properties are listed in Table 4. Due to the high Li concentration, the Li-Li repulsive interactions play the role and compete with the Li-P attractive interactions in stabilizing the systems. For example, in the initial configuration of S-4(VH-VH), S-4(VH-VB)$_{a/b}$, S-4(VH-TB)$_{a/b}$, the distances of Li-Li pairs were 0.86 Å, 1.71 Å, and 2.36 Å, respectively, which are much shorter than the Li-Li bond length of 2.68 Å. During the relaxation, Li atoms at the VH sites will push Li atoms at other sites in the same unit cell to the positions at TH sites. As the results, these initial configurations were finally stabilized to the structures with S-4(VH-TH) configuration (see the first panel in Fig. 9 (a)) with the Li-Li distances increased to 3.0 Å (about 0.3 Å larger than the Li-Li bond length of 2.68 Å, see the 6$^{th}$ column in Table 4). Such system has the lowest adsorption energy (-1.826 eV/Li) among other single-side configurations. Another example is found in the initial configurations S-4(VB-TB)$_{a/b}$, in which the initial Li-Li distance is either 2.90 Å (for type a) or 2.39 Å (for type b). During the relaxation, Li atoms at TB sites were pushed to a different vertical level (*e.g.*, see the 2$^{nd}$ and 3$^{rd}$ panels in Fig. 9 (a)) with the Li-Li distances increased to 2.92-3.02 Å. Similar results are also found in the initial configurations S-4(VB-VB) and S-4(TB-TB), in which the initial Li-Li distances were set around 1.65 Å. During the relaxation, we found that two Li atoms in each unit cell push each other vertically to reduce the Li-Li repulsive interactions, resulting one of the Li atoms moved up away from phosphorene layer with the vertical distance of 5.42/6.02 Å and the Li-Li distance of 3.03/3.06 Å. While, these Li atoms were still on the top of the VB or TB site with the same type of configurations as S-4(VB-VB) and S-4(TB-TB) (see the last two panels in Fig. 9 (a)).

The double-side configurations for $Li_8P_{16}$ system were constructed by adding another 4 more Li atoms on the other side of the stable single-side $Li_4P_{16}$ systems. Thus, in each 1x1 unit cell, one Li atom is located at the VH, or the VB or TB site above phosphorene layer and the other is located at the VH, or VB or TB site below phosphorene layer. Obtained the stable double-side configurations of $Li_8P_{16}$ system are shown in Fig. 9 (b), named as D-4(VH/VH), D-4(VH/VB), D-4(VB/VB)$_{a/b}$, D-4(VB/TB)$_{a/b}$, and D-4(TB/TB)$_{a/b}$, respectively, where subscriptions a and b distinguish the different symmetry in the same configuration. The corresponding adsorption energies and geometric properties are also listed in Table 4. As can be seen from Table 4, the most stable structure among the stable single/double-side configurations in $Li_8P_{16}$ system is the one with D-4(VH/VH) configuration ($E_a$=-2.007 eV/Li). Another three double-side configurations (*i.e.*, D-4(VH/VB) and D-4(VB/VB)$_{a/b}$) are about 0.042-0.085 eV higher than the configuration of D-4(VH/VH), but about 0.096-0.16 eV lower than the two single-side configurations (*i.e.*, S-4(VH-TH) and S-4(VB-TB)$_a$). Furthermore, the single-side configuration (*i.e.*, S-4(VB-TB)$_b$) with four Li atoms pulled up away from the phosphorene layer is still competitive with the double-side configurations of D-4(VB/TB)$_a$ and D-4(VB/TB)$_b$. More interesting is found that the single-side configuration S-4(VB/VB) is even relatively stable than the double-side configuration D-4(TB/TB)$_{a/b}$, indicating that when more Li atoms load on the phosphorene, some Li atoms prefer to stay at single-side at VB sites with higher vertical distance instead of staying at double sides at TB sites with lower vertical distances. Comparing the nearest Li-P distance and vertical distance $d$ with the corresponding double-side Li$_4$P$_{16}$ configurations, we found that the Li-P distance does not change too much (see the 4$^{th}$ column in Table 4), but the vertical distance $d$ becomes higher (see the 5$^{th}$ column in Table 4), indicating a small vertical elongation of phosphorene layer due to the more Li-P attractions when more Li atoms were added on the phosphorene. Again, in all stabilized structures of the Li$_8$P$_{16}$ system, the repulsive force between Li atoms drove Li atoms to occupy different sites with Li-Li distance around 3.0 Å (see the 6$^{th}$ column in Table 4), and no Li-clustering was found.

### 4.3.5 Li$_{16}$P$_{16}$ system

At this high Li concentration, we focused on finding the possible existence of double-side configurations in Li$_{16}$P$_{16}$ system with no Li clustering. Follow the same scheme that we used in constructing double-side configurations for Li$_n$P$_{16}$ ($n$=2, 4, 8), we added 8 Li atoms on the other side of the five stable single-side configurations obtained in Li$_8$P$_{16}$ system to construct the double-side configurations for Li$_{16}$P$_{16}$, and 5 stable structures are obtained after fully relaxation. They are named by D-4(VH-VH/VH-VH), D-4(VB-TB/VB-TB)$_{a/b}$, D-4(VB-VB/VB-VB), and D-4(TB-TB/TB-TB), respectively. Fig. 10 shows the top and side views of these stable structures, and Table 5 lists the corresponding adsorption energies and geometric properties. At this high Li concentration, we found two cases of Li atoms adsorbed on phosphorene layer. In the first case, due to the strong Li-P attractive interaction between Li and P atoms on both sides, the puckered phosphorene layer was separated during the relaxation, forming buckled up and down zigzag chains (see the configurations D-4(VH-VH/VH-VH), D-4(VB-TB/VB-TB)$_a$ and D-4(VB-TB/VB-TB)$_b$ in Fig. 10). In the second case, half of the Li atoms located at the zigzag valley move up/down from phosphorene layer (see the configurations D-4(VB-VB/VB-VB), and D-4(TB-TB/TB-TB) in Fig. 10). Interesting is found that the structures in the first case are relatively stable than these in the second case (see the 3$^{rd}$ column in Table 5), in particular, the structure with all Li atoms in the zigzag valley (*i.e.*, the D-4(VH-VH/VH-VH) configuration) is the most stable (*i.e.*, $E_b$=-2.046 eV/Li) in the Li$_{16}$P$_{16}$ system, and even more stable than the most stable structure in Li$_8$P$_{16}$ system (*i.e.*, the D-4(VH/VH) configuration), indicating the ability of phosphorene to accommodate four Li atom per 1x1 unit cell. But the cost is the volume expansion/bond broken between the P atoms. We found that, during the relaxation process, 8 Li atoms on each side of phosphorene, for example in D-4(VH-VH/VH-VH) configuration (see Fig. 11 (a)), strongly attract P atoms on both side of phosphorene layer, leading to the opposite motion of P atoms. As the result, the P-P bonds between P atoms at the valley and ridge were broken, and the puckered phosphorene layer was separated to two buckled above/below zigzag chains with the separation of about 3.61 Å, similar as a volume expansion of 56% during the charging process in LIB. So, the question is whether such broken bonds could be self-reformed after Li atoms removed from phosphorene? To unravel this puzzle, we examined the Li$_{16}$P$_{16}$ structures in the first case by relaxing the systems after removing Li atoms.

Fig. 11 (b) shows, for example, the relaxation process of the stable Li$_{16}$P$_{16}$ system with the configuration D-4(VH-VH/VH-VH) after removing the 16 Li atoms (mimic the delithiation process). Since all Li atoms are removed, there is no Li-P attractive force to pull phosphorus away from each other, and the buckled zigzag chains became flat first (in about 17 fs) and then moved close each other reforming the P-P bonds, and finally, the puckered phosphorene structure was recovered (in about 10 fs). The volume expansion of the phosphorene during Li insertion/discharging disappeared after the Li extraction, showing the reversibility of the phosphorene during Li insertion/extraction process as an anode material. We should point out that even though the Li atoms are close each other at this high Li concentration, and the Li-Li distance is about 0.05-0.1 Å shorter than in the other $Li_nP_{16}$ systems ($n$=2, 4, 8) (see the last column in Table 5), there is still no Li cluster formed. In fact, the adsorption energies in all the stable configurations at this Li high concentration were lower than the cohesive energy of Li metal (~ 1.63 eV[94]), which ensure to prevent the Li clustering during Li intercalation.

### 4.4 Electronic properties of $Li_nP_{16}$ systems

The electronic properties of phosphorene with various $Li_nP_{16}$ configurations were studied from the total and projected density of states (*i.e.*, DOS and PDOS). The DOS/PDOS for the four most stable single-side configurations in each $Li_nP_{16}$ system ($n$=1, 2, 4, 8) and the four stable double-side configurations in each $Li_nP_{16}$ system ($n$=2, 4, 8, 16) were presented in Fig. 12 (a) and (b), respectively. The DOS of pristine phosphorene with the DFT band gap of 0.82 eV was also shown on the top of Fig. 12 (a) and (b).

It can be seen from Fig. 12 (a) and (b) that during the Li insertion, the tail at the bottom of the conduction band (CB) extends towards to the top of the valence band (VB) due to the orbital hybridization between Li 2s and P 2p orbitals. On the other hand, the fermi level upshifts to the CB, due to the charge transfer from Li to the phosphorene, indicating the ionic bonding between Li and P. The PDOS of Li are mainly appear in the CB near the fermi energy, indicating that the 2s electron of Li contribute to the DOS near the fermi level. With more Li atoms adsorbed, more active charge transfer from the Li atoms to the phosphorene leading to the fermi level shift even further, more Li PDOS appear close to the bottom of CB, and the energy band gap gradually diminishes and eventually disappears. In particular, the semiconductor-metal transition was observed in both single-side and double-side adsorption which is expected for the electron transport to be sufficiently fast in the phosphorene anode.

**4.5 Li capacity of phosphorene**

The summary of adsorption energy as a function of Li concentration was shown in Fig. 13. The black up-triangles indicate the adsorption energy of the $Li_nP_{16}$ systems with single-side adsorption, and the red down-triangles indicate the adsorption energy of the $Li_nP_{16}$ systems with double-side adsorption, respectively. All negative adsorption energy values indicate the stability of the systems. It also indicates that the weak van der Waals interactions might play a minor role in the Li adsorption and could be negligible, as we assumed during our DFT calculations. As discussed in section 4.4, among all the obtained stable structures with the Li adsorption either at the single-side or at the double-side, the most stable configurations for each $Li_nP_{16}$ system (*i.e.*, up/down triangles at the bottom guided with the black dashed and red dash-dotted curves in Fig. 13) are those in which Li atoms always prefer to stay at VH sites along the zigzag direction in the valley. This makes sense because these VH sites are the most preferential sites as predicted from the adsorption energy landscape (as shown in Fig. 3). On the other hand, in the $Li_nP_{16}$ systems with higher adsorption energies (*e.g.*, the up/down triangles above the black dashed and the red dash-dotted curves in Fig. 13) more Li atoms stay at VB and TB sites. It was found that for the most table single-side configurations in Li$_n$P$_{16}$ systems, the adsorption energy increases with increasing the number of Li atoms (guided by the black dashed curve in Fig. 13), while, the adsorption energy for the most stable double-side configurations increases first at small Li concentration (0<*n*<8), reaches the maximum at *n*=8, and then decreases after *n*>8 (guided by the red dash-dotted curve for the double-side adsorption in Fig. 13). In particular, the adsorption energy in the most table double-side configurations, for given *n* in Li$_n$P$_{16}$ systems, is always lower than that in the most stable single-side configurations, as the repulsive Li-Li interaction is weaker in the former than in the latter. The theoretical specific capacity was evaluated from Eq. (3). Since the maximum Li to P ratios are 0.5 (Li$_8$P$_{16}$) for single-ide adsorption and 1.0 (Li$_{16}$P$_{16}$) for double-side adsorption, respectively, the estimated corresponding specific capacity for phosphorene as an anode is predicted as 433 *mAh/g* single-ide adsorption and 865 *mAh/g* for double-side adsorption, respectively.

We performed the very first experimental measurement of lithium storage capacity of few layer phosphorene networks. Fig. 14 shows the galvanostatic charge-discharge curves of phosphorene/TAB-2 electrode for rechargeable Li-ion battery at the voltage range of 0.05-2.8 V at C/10 rate. It has an initial irreversible discharge capacity of 3065 *mAh/g* during the first cycle. The discharge potential plateau curve was large and flat at ~1 V and 0.4 V vs. Li/Li+. Furthermore, charge–discharge measurements of the phosphorene/TAB-2 (90/10) electrode show the discharge capacities of 685, 477, and 453 *mAh/g* during the 2$^{nd}$, 10$^{th}$, and 30$^{th}$ cycles, respectively. After 30 cycles, it retains very stable discharge capacity of 453 *mAh/g*. The first discharge capacity (Li-insertion) usually is gained 3065 *mAh/g* due to the reaction of the electrolyte at the surface of phosphorene with transferred lithium atoms to form a passivating film named as solid electrolyte interface (SEI) process. Once, SEI is formed, it prevents the further electrolyte reaction on the phosphorene surface. Thus, the first discharge profile is always different from the profiles of subsequent cycles. Nevertheless, the capacity is stabilized after $30^{th}$ cycle and remained highly reversible over 50 cycles. It is expected that the reversible capacity can reach the theoretically predicted value in case if the single layered phosphorene can be arranged as 'house card' as found in the case of graphene [43], or capped with other 2D materials [94].

## 5. CONCLUSION

The novel features of phosphorene as anode materials for LIB have been characterized based on the first principle calculations and experimental measurements. When adsorbed on phosphorene, Li atoms prefer to reside at the most stable VH sites at the valley. The other two metastable sites (*i.e*., the VB and TB sites) could be occupied when there more Li atoms loaded on the phosphorene. The adsorption energy at the most stable VH site is -2.086 eV/Li, indicating strong Coulomb interactions between Li and phosphorene and the ability of having high open circuit voltage (~ 2.5 V), which are essential in the electrochemical performance for high performance LIB. The diffusion energy barrier of Li on phosphorene shows high anisotropic behavior and is extremely low when Li atom migrates along the zigzag channel at the valley. Estimated diffusion constant also shows that Li atoms will diffuse ultrafast and directionally along the zigzag direction in the valley (*e.g*., about $10^{11}$ times faster than the diffusion across the ridge along the armchair direction). In particular, it was found that the Li diffusion on phosphorene is extremely faster than on the graphene and $MoS_2$, which implies that the phosphorene may exhibit outstanding high rate capacity.

The most stable structures for Li atoms adsorbed on single/double-side of the phosphorene layer are those in which Li atoms occupy at the VH sites along the zigzag direction in the valley. Importantly, the ability for phosphorene to accommodate Li atoms was found up to about 1:1 (*i.e.*, the ratio of Li:P), demonstrating that the monolayer phosphorene could reach the theoretical capacity with 865 *mAh/g*. Experimental measured specific capacity for a few layered phosphorene network showed very stable value of 453 after $30^{th}$ cycle and good cycling performance. A uniform single layered phosphorene with nanoporosity is expected to increase the reversible capacity to the theoretically predicted value. More interesting was found that phosphorene monolayer could self-recover when it was 'distorted/fractured' during Li intercalation at the high Li:P ratio, indicating its reversibility during lithiation/dilithiation (charging/discharging) process. Overall, our theoretical and experimental results show the beneficial properties of Li adsorbed phosphorene including the high specific capacity, the ultrafast and anisotropic diffusivity, reversibility in charging/discharging, stable cycling performance, high OCV, and electrochemical performance, which make it an excellent candidate as anode material for high performance Li-ion batteries.

ACKNOWLEDGMENT: C. Zhang and M. Yu acknowledge computing resource support from the Cardinal Research Cluster at the University of Louisville. G. Sumanasekera acknowledges the support by the KY NSF-EPSCOR under grant no. T1 2014-2019. G. Anderson acknowledges the support received from the Mc Sweeny Fellowship of the College of Arts and Sciences at the University of Louisville.

Table 1 Calculated adsorption energy ($E_a$), the distances between Li and the nearest P atoms on phosphorene ($d_{Li-P}$), and the vertical distance ($d$) of LiP$_{16}$ system with Li atom at the preferential adsorption positions (VH, VB, and TB sites).

| System | Adsorption Site | $E_a$ (eV/Li) | $d_{Li-P}$ (Å) | $d$ (Å) |
|---|---|---|---|---|
| | VH | -2.086 | 2.45, 2.54 | 2.53 |
| LiP$_{16}$ | VB | -1.995 | 2.41 | 2.61 |
| | TB | -1.427 | 2.58 | 3.23 |

Table 2 Calculated adsorption energy ($E_a$), distances between Li and the nearest P atoms on phosphorene ($d_{Li-P}$), the vertical distance ($d$), and the nearest Li-Li distance ($d_{Li-Li}$) for the stable single-side (up panel) and double-side (bottom panel) configurations of the $Li_2P_{16}$ system. In case there are different types of $d_{Li-P}$ and $d$ values for a given configuration, notations corresponding to the sites are indicated in the parentheses in columns 4 and 5. For example, in the configuration D-VH/VB, the $d_{Li-P}$ and $d$ values at VH and VB sites are indicated by VH and VB in the corresponding parentheses. Note that the large Li-Li distances (beyond 6.5 Å) in the double-side configurations in the $Li_2P_{16}$ system are not counted here.

| System | Configuration | $E_a$ (eV/Li) | $d_{Li-P}$ (Å) | $d$ (Å) | $d_{Li-Li}$ (Å) |
|---|---|---|---|---|---|
| $Li_2P_{16}$ (Single-side) | S-VH-VH | -2.045 | 2.35, 2.56 | 2.54 | 4.62 |
|  | S-VH-VB | -2.000 | 2.38, 2.53(VH) 2.32(VB) | 2.54(VH) 2.63 (VB) | 4.99 |
|  | S-VB-VB | -1.957 | 2.31 | 2.63 | 4.62 |
|  | S-TB-TB | -1.347 | 2.45 | 3.28 | 4.62 |
| $Li_2P_{16}$ (Double-side) | D-VH/VH | -2.113 | 2.43, 2.53 | 2.51 | — |
|  | D-VH/VB | -2.064 | 2.43, 2.53 (VH) 2.41 (VB) | 2.50(VH) 2.58 (VB) | — |
|  | D-VB/VB | -2.018 | 2.41 | 2.60 | — |
|  | D-VH/TB | -1.755 | 2.34, 2.55 (VH) 2.50 (VB) | 2.53(VH) 3.23(TB) | — |
|  | D-VB/TB | -1.718 | 2.40(VB), 2.47(TB) | 2.59 (VB) 3.23 (TB) | — |

Table 3 Calculated adsorption energy ($E_a$), distances between Li and the nearest P atoms on phosphorene ($d_{Li-P}$), the vertical distance ($d$), and the nearest Li-Li distance ($d_{Li-Li}$) for the stable single-side (up panel) and double-side (bottom panel) configurations of the $Li_4P_{16}$ system. Note that in the configuration D-(VH-VB)/(VH-VB), the $d_{Li-P}$ and $d$ values at VH and VB sites are indicated by VH and VB in the corresponding parentheses, respectively.

| System | Configuration | $E_a$ (eV/Li) | $d_{Li-P}$ (Å) | $d$ (Å) | $d_{Li-Li}$ (Å) |
|---|---|---|---|---|---|
| $Li_4P_{16}$ (Single-side) | S-4VH$_a$ | -1.968 | 2.46, 2.53 | 2.64 | 3.31 |
| | S-4VH$_b$ | -1.949 | 2.42, 2.56 | 2.64 | 3.31 |
| | S-4VB | -1.885 | 2.41 | 2.74 | 3.31 |
| | S-2(VH-TH) | -1.769 | 2.44, 2.63(VH) / 2.69 (TH) | 2.75(VH) / 3.58(TH) | 3.00 |
| | S-4TB | -1.404 | 2.58 | 3.41 | 3.31 |
| $Li_4P_{16}$ (Double-side) | D-2(VH/VH)$_a$ | -2.090 | 2.41, 2.51 | 2.52 | 4.62 |
| | D-2(VB/VB) | -2.073 | 2.31 | 2.64 | 4.62 |
| | D-(VH-VB)/(VH-VB) | -2.048 | 2.41, 2.51 (VH) / 2.40 (VB) | 2.51(VH) / 2.60 (VB) | 4.40 |
| | D-2(VH/VH)$_b$ | -1.978 | 2.41, 2.54 | 2.48 | 3.31 |

Table 4 Calculated adsorption energy ($E_a$), distances between Li and the nearest P atoms on phosphorene ($d_{Li-P}$), the vertical distance ($d$), and the nearest Li-Li distance ($d_{Li-Li}$) for the stable single-side (up panel) and double-side (bottom panel) configurations of the $Li_8P_{16}$ system. In case there are different types of $d_{Li-P}$ and $d$ values for a given configuration, notations

corresponding to the sites are indicated in the parentheses in columns 4 and 5. For example, in the configuration S-4(VH/TH), the $d_{Li-P}$ and $d$ values at VH and TH sites are indicated by VH and TH in the corresponding parentheses, and in the configuration S-4(VB-VB), the $d_{Li-P}$ and $d$ values at VB sites up and down positions are indicated by up and down in the corresponding parentheses, respectively.

| System | Configuration | $E_a$ (eV/Li) | $d_{Li-P}$ (Å) | $d$ (Å) | $d_{Li-Li}$ (Å) |
|---|---|---|---|---|---|
| Li$_8$P$_{16}$ (Single-side) | S-4(VH-TH) | -1.826 | 2.43, 2.59(VH) <br> 2.78 (TH) | 2.69 (VH) <br> 3.68 (TH) | 3.00 |
| | S-4(VB-TB)$_a$ | -1.805 | 2.44(VB) <br> 2.94 (TB) | 2.78 (VB) <br> 3.80 (TB) | 3.02 |
| | S-4(VB-TB)$_b$ | -1.712 | 2.46 (VB) <br> 3.70 (TB) | 2.81 (VB) <br> 4.60 (TB) | 2.92 |
| | S-4(VB-VB) | -1.658 | 2.49 | 2.84 (down) <br> 4.60 (up) | 3.06 |
| | S-4(TB-TB) | -1.481 | 2.65 | 3.48 (down) <br> 6.02 (up) | 3.03 |
| Li$_8$P$_{16}$ (Double-side) | D-4(VH/VH) | -2.007 | 2.40, 2.53 | 2.62 | 3.31 |
| | D-4(VH/VB) | -1.965 | 2.41, 2.54(VH) <br> 2.40(VB) | 2.63(VH) <br> 2.74 (VB) | 3.31 |
| | D-4(VB/VB)$_a$ | -1.928 | 2.40 | 2.74 | 3.31 |
| | D-4(VB/VB)$_b$ | -1.922 | 2.40 | 2.75 | 3.31 |
| | D-4(VB/TB)$_a$ | -1.719 | 2.40(VB) <br> 2.53(TB) | 2.78 (VB) <br> 3.44(TB) | 3.31 |
| | D-4(VB/TB)$_b$ | -1.716 | 2.40 (VB) <br> 2.53 (TB) | 2.78(VB) <br> 3.44(TB) | 3.31 |

| | | | | |
|---|---|---|---|---|
| D-4(TB/TB)$_a$ | -1.520 | 2.53 | 3.43 | 3.31 |
| D-4(TB/TB)$_b$ | -1.517 | 2.53 | 3.43 | 3.31 |

Table 5. Calculated adsorption energy ($E_a$), distances between Li and the nearest P atoms on phosphorene ($d_{Li-P}$), the vertical distance ($d$), and the nearest Li-Li distance ($d_{Li-Li}$) for double-side configurations of the Li$_{16}$P$_{16}$ system. In case there are different types of $d_{Li-P}$ and $d$ values for a given configuration, notations corresponding to the sites are indicated in the parentheses in columns 4 and 5. For example, in the configuration D-4(VB-TB/VB-TB)$_a$, the $d_{Li-P}$ and $d$ values at VB and TB sites are indicated by VB and TB in the corresponding parentheses.

| System | Configuration | $E_a$ (eV/Li) | $d_{Li-P}$ (Å) | $d$ (Å) | $d_{Li-Li}$ (Å) |
|---|---|---|---|---|---|
| Li$_{16}$P$_{16}$ (Double-side) | D-4(VH-VH/VH-VH) | -2.046 | 2.38, 2.50, 2.58 | 3.30, 4.17 | 2.91 |
| | D-4(VB-TB/VB-TB)$_a$ | -1.896 | 2.40(TB) 2.64(VB) | 2.82 (TB) 3.88 (VB) | 2.95 |
| | D-4(VB-TB/VB-TB)$_b$ | -1.757 | 2.34(TB) 2.79(VB) | 2.93(TB) 3.74(VB) | 2.55 |
| | D-4(VB-VB/VB-VB) | -1.752 | 2.50 | 2.84, 5.43 | 3.306 |
| | D-4(TB-TB/TB-TB) | -1.648 | 2.65 | 3.48, 6.03 | 3.306 |

FIGURE CAPTIONS

Figure 1 (color online) (a) Raman spectrum of phosphorene obtained from different areas of sample vial after centrifugation, bulk black phosphorus, and bulk red phosphorus. (b) Photoluminescence of phosphorene obtained from different areas of sample vial after centrifugation and bulk black phosphorus. (c) AFM image of monolayer phosphorene. (d) SEM image of overlapping phosphorene flakes.

Figure 2 (color online) (a) Top view of phosphorene with 40 single Li adsorption sites in the 1x1 unit cell (*i.e.*, the black dashed box). The larger balls with light (dark) purple color denote phosphorus on the ridge (in the valley). The smaller balls with different color indicate positions of Li atoms, in which the red color represents the position of a Li on the top of P atoms (*i.e.*, the TA sites); the blue color, on the top of P-P bonds (*i.e.*, VB and TB sites), the green color, at the center of triangular region formed by P-P-P atoms (*i.e.*, VH and TH sites); and the black color, at other sites of unit cell above the phosphorene. (b) The vertical distance $d$ defined from Li atom (black ball) to the middle of puckered phosphorene indicated by the dot-dash line. (c) The total energy as a function of the vertical distance $d$ for a Li atom at the VH site.

Figure 3 (color online) (a) Adsorption energy landscape of a Li atom on phosphorene. The colors in the right column represent the adsorption energy at an adsorption site relative to that at the VH site. The darker the color is the lower in relative adsorption energy. A schematic structure of the 2×2 phosphorene supercell is attached on the landscape to illustrate the Li adsorption positions. (b) and (c) zoom in the energy landscape in the white box (around the saddle points at VB site) and blue box (around the saddle points at TB site) in (a), respectively.

Figure 4 (color online) (a) Top view of the diffusion pathways on monolayer phosphorene (see the definition for each pathway in the text). Black dots denote the path 1 (VH→VB→VH); green dots, the path 2 (VB→TB→VB); blue dots, the path 3 (VH→TB→VH); and red dots, the path 4 (VH→TA→VH), respectively. (b) and (c) are the side views seen from the bottom and the right side of (a), respectively.

Figure 5 (color online) (a) The energy barrier (see the definition in the text) as a function of the relative distance along the four paths (indicated in Fig. 3 (a)). The relative distance is defined by the ratio of the horizontal distance from the starting site to the position of the Li atom along a given pathway and that from the starting site to the ending site. (b) The vertical distance $d$ as a function of the relative distance. The corresponding top and side views of these four paths are shown in Fig. 3.

Figure 6 (color online) The top (up) and side (down) views of the stable $LiP_{16}$ configurations with the Li atom adsorbed at the VH site (left), the VB site (middle), and the TB site (right), respectively. The dark gray ball represents the Li atom, and the balls with light and dark purple colors denote phosphorus atoms on the ridge and in the valley, respectively. Please note that, in order to see the distribution of Li atoms on phosphorene clearly, Li-P bonds were not presented here, the same reason for the following Figures 7-10.

Figure 7 (color online) The top (up) and side (down) views of the stable $Li_2P_{16}$ system with single-side configurations (a) and double-side configurations (b), respectively. The light and dark gray balls represent Li atoms above and below the phosphorene monolayer, respectively; while, the balls with light and dark purple color denote phosphorus on the ridge and in the valley, respectively. For the notations corresponding to each configuration, please see their definitions in the text.

Figure 8 (color online) The top (up) and side (down) views of the stable $Li_4P_{16}$ system with single-side configurations (a) and double-side configurations (b). The light and dark gray balls represent Li atoms above and below the phosphorene monolayer, respectively; while, the balls with light and dark purple color denote phosphorus on the ridge and in the valley, respectively. For the notations corresponding to each configuration, please see their definitions in the text.

Figure 9 (color online) The top (up) and side (down) views of the stable $Li_8P_{16}$ system with single-side configurations (a) and double-side configurations (b). The light and dark gray balls represent Li atoms above and below the phosphorene monolayer, respectively; while, the balls with light and dark purple color denote phosphorus on the ridge and in the valley, respectively. For the notations corresponding to each configuration, please see their definitions in the text.

Figure 10 (color online) Top (up) and side (down) views of the stable $Li_{16}P_{16}$ system with double-side configurations. The light and dark gray balls represent Li atoms above and below the phosphorene monolayer, respectively; while, the balls with light and dark purple color denote phosphorus on the ridge and in the valley, respectively. For the notations corresponding to each configuration, please see their definitions in the text.

Figure 11 (color online) The relative energy as function of molecular dynamics (MD) steps for (a) the relaxation procedure of 16 Li atoms inserted to both sides of phosphorene layer with the initial configuration D-4(VH-VH/VH-VH) and (b) the relaxation procedure of the 'fractured' phosphorene layer which was obtained by removing the 16 Li atom from the fully relaxed $Li_{16}P_{16}$ system with the configuration D-4(VH-VH/VH-VH) (*i.e.,* the fully relaxed structure showed in the case (a)). The relative energy is defined as the total energy difference between the initial and final stage of the system. The time step was set 0.5 fs in the MD simulation. The top (up) and side (down) views of several intermediate configurations are illustrated in the insets with the starts representing the relative energy at corresponding MD steps. The bond reforming could be seen after the 55 MD steps (see (b)). The light and dark gray balls represent Li atoms above and below the phosphorene monolayer, respectively; while, the balls with light and dark purple color denote phosphorus on the ridge and in the valley, respectively.

Figure 12 (color online) The total (DOS) and partial (PDOS) densities of states of some stable single-side (a) and double-side (b) configurations of $Li_nP_{16}$ system with $n$=1, 2, 4, 8, and 16, respectively. The Total DOS are plotted with black solid curves, and the partial DOS for P (Li) atoms are plotted with red dot-dash (blue dash) curves, respectively. The fermi level is represented by black dash line. The DOS for the pristine phosphorene is also shown on the top of (a) and (b).

Figure 13 (color online) The adsorption energy per Li atom of the $Li_nP_{16}$ system with various stable configurations versus the number of Li atoms $n$. The black up triangles and the red down triangles represent the system with the single-side and the double-side adsorptions, respectively. The black dash curve (the red dot-dash curve) was used to guide the lowest adsorption energy of the stable $Li_nP_{16}$ system with single-side (double-side) adsorption for given $n$.

Figure 14 (color online) Electrochemical characterization of Phosphorene-based coin cells. (a) Charge-discharge Voltage versus capacity curves tested at C/10 rate plotted after the 1st (black), 2nd (pink), 10th (blue), 20$^{th}$ (green), and 50th (red) cycles. (b) Specific discharge capacity cycling at C/10 rate.

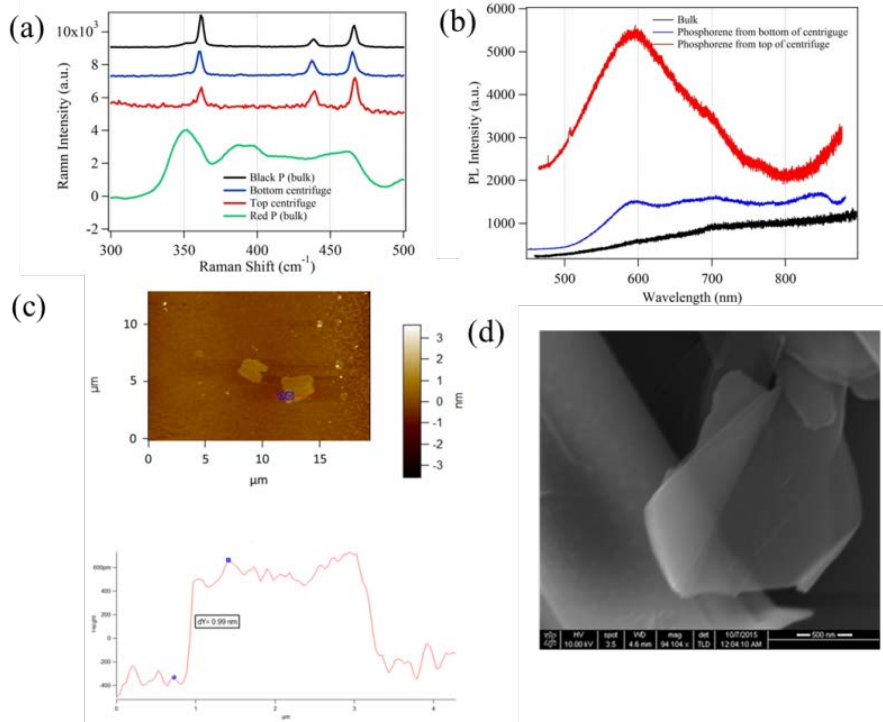

Fig. 1

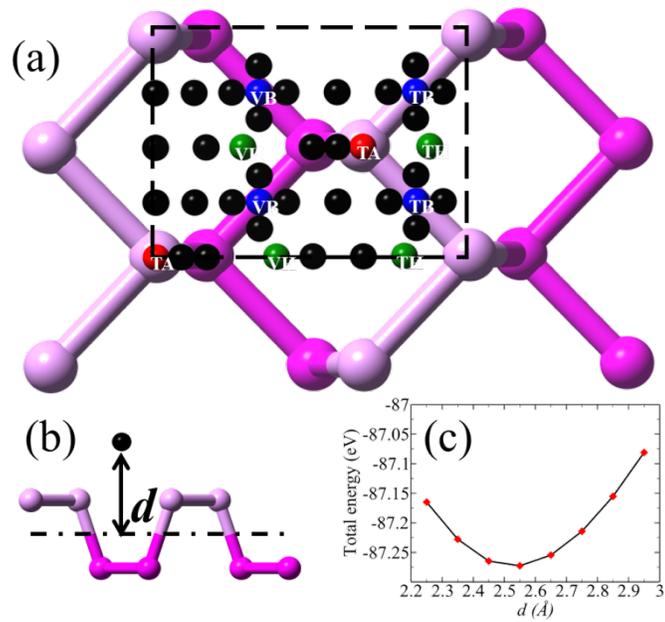

Fig. 2

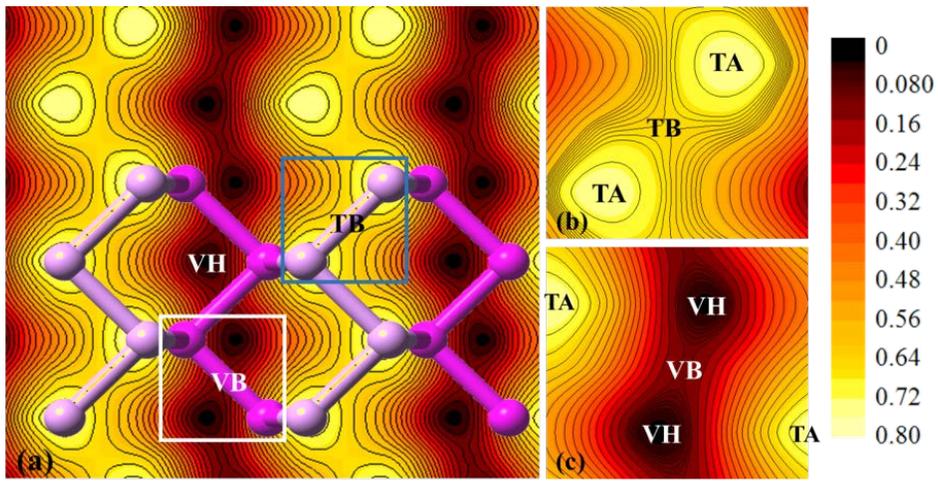

Fig. 3

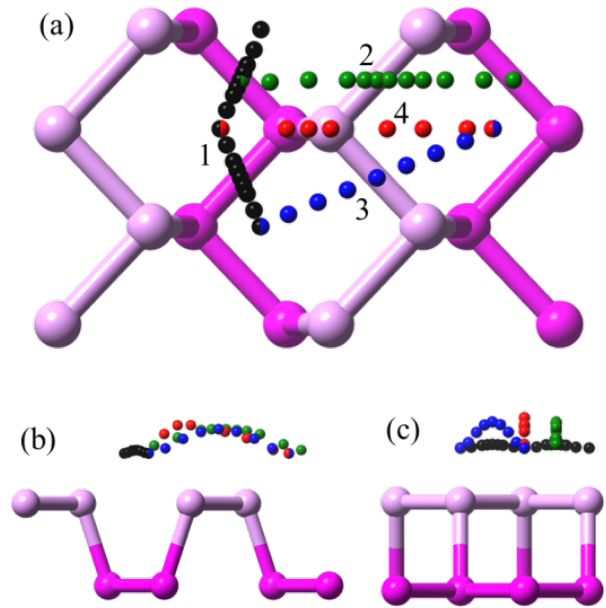

Fig. 4

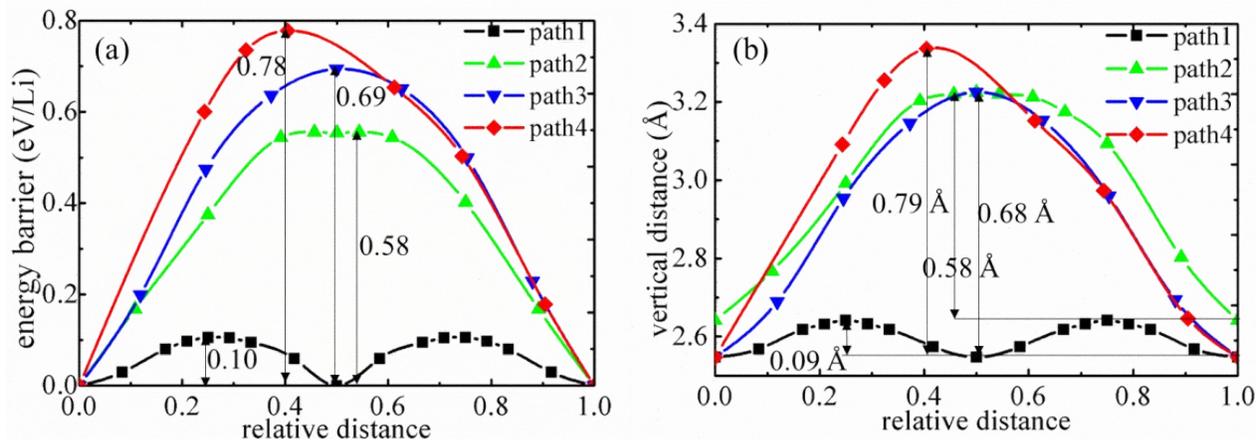

Fig. 5

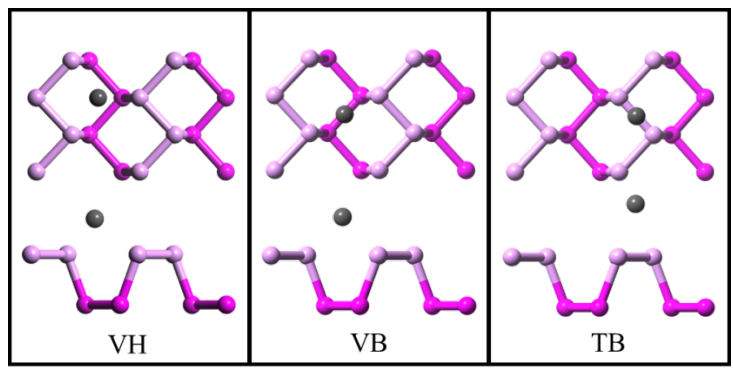

Fig. 6

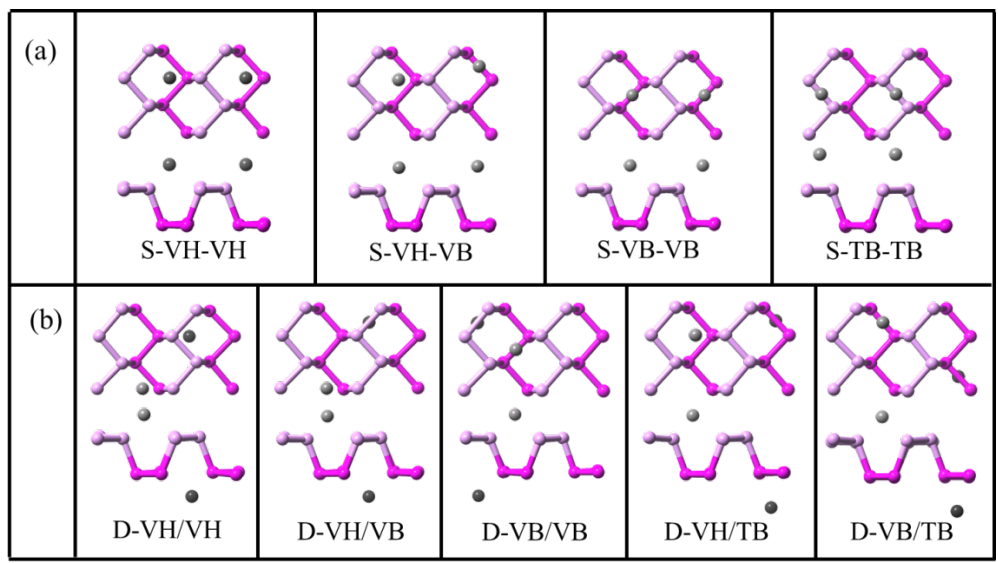

Fig. 7

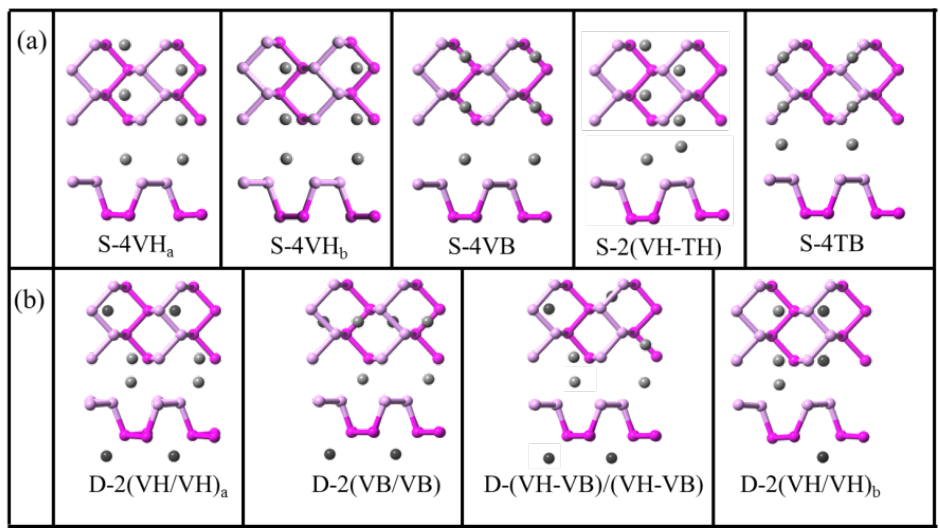

Fig. 8

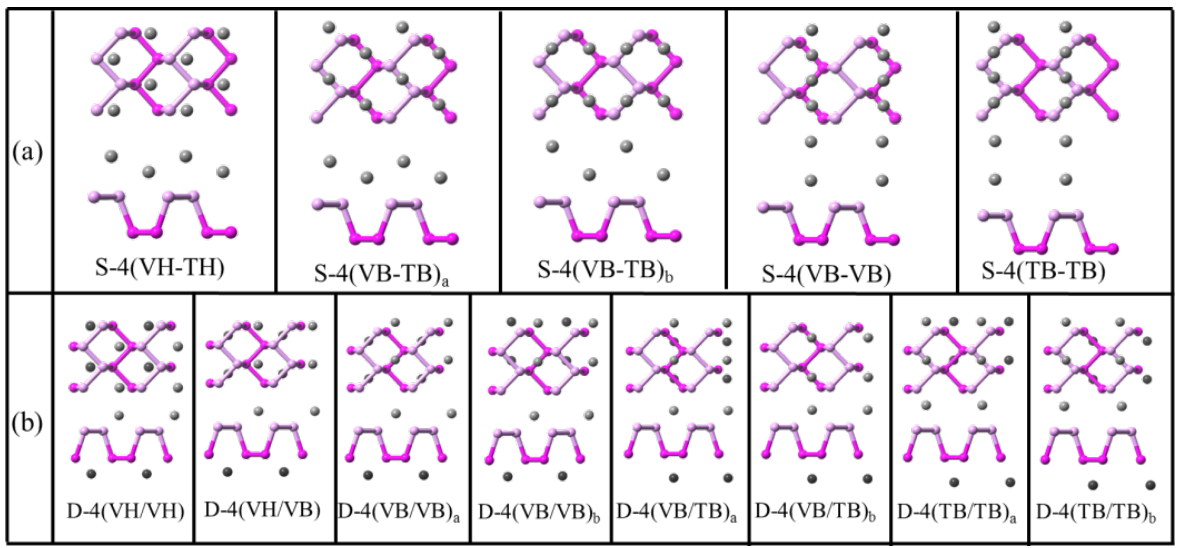

Fig. 9

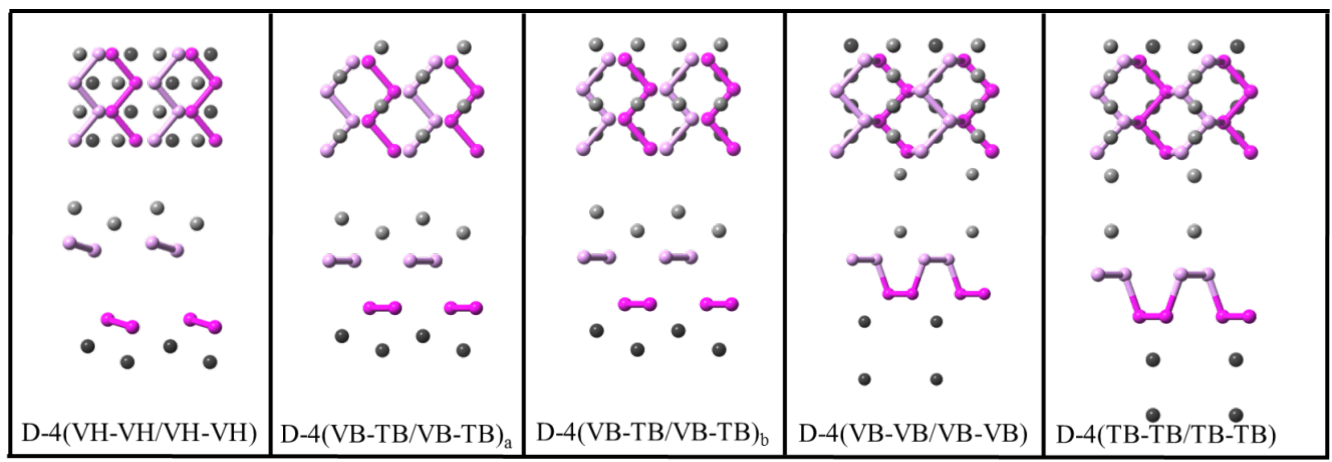

Fig. 10

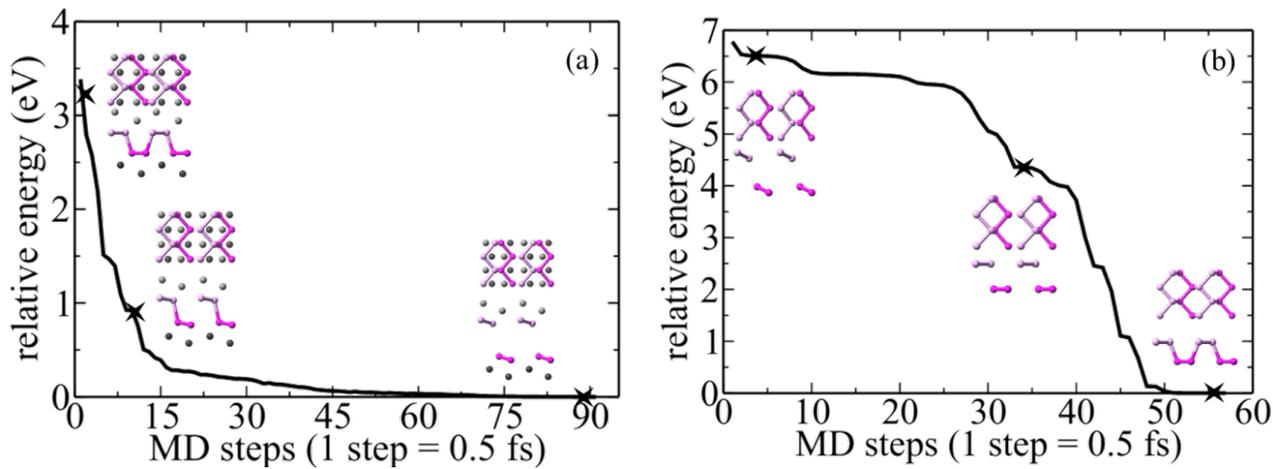

Fig. 11

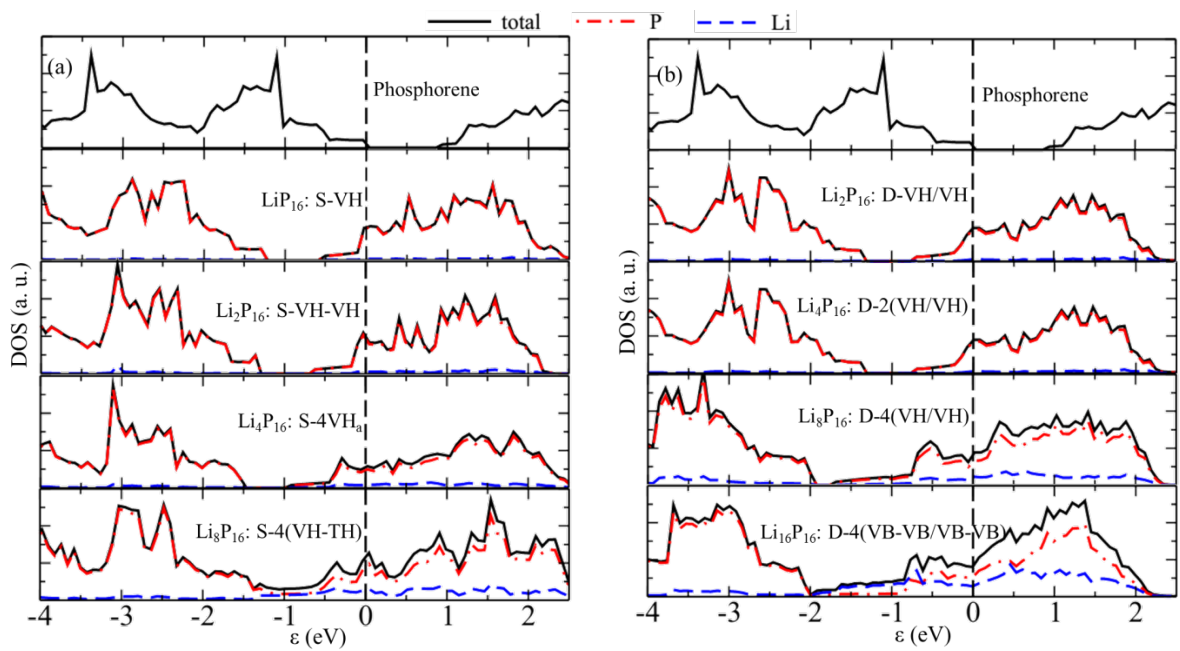

Fig. 12

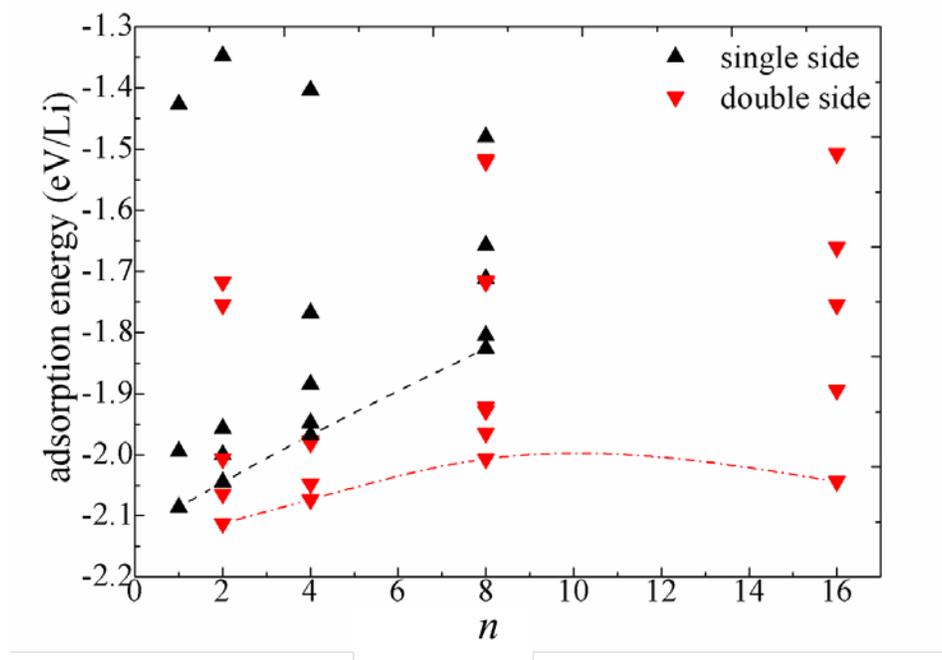

Fig. 13

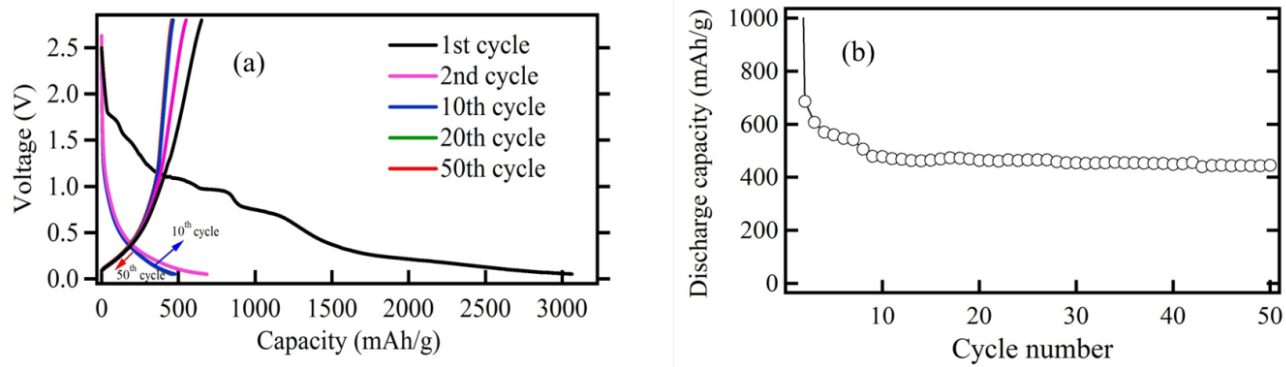

Fig. 14